\documentclass[traditabstract]{aa} 

\usepackage{txfonts} 
\usepackage{multirow,longtable,psfig}

\usepackage{graphicx}

\newcommand \sw{{\it Swift}}
\newcommand \ba{{ BATSE}}
\newcommand \fe{{\it Fermi}}

\newcommand \epo{$E^{\rm obs}_{\rm peak}$}

\newcommand \gsim{ \lower .75ex \hbox{$\sim$} \llap{\raise .27ex \hbox{$>$}} } 
\newcommand \lsim{ \lower .75ex\hbox{$\sim$} \llap{\raise .27ex \hbox{$<$}} }

\begin{document}
\title{Spectral properties of 438 GRBs detected by \fe /GBM}

\author{L. Nava \inst{1}\thanks{lara.nava@sissa.it}
 \and G. Ghirlanda \inst{2} \and G. Ghisellini \inst{2} \and A. Celotti \inst{1}}
\institute{SISSA, via Bonomea 265, I--34136 Trieste, Italy
\and Osservatorio Astronomico di Brera, via E. Bianchi 46, I--23807 Merate, Italy }
\date{Received .. ... .. / Accepted .. ... ..} 

\abstract{
We present the results of the spectral analysis of the public data of 438 Gamma Ray Bursts (GRBs) 
detected by the \fe\ Gamma ray Burst Monitor (GBM) up to March 2010. For 432 bursts we could fit the time integrated spectrum. 
In 318 cases we can reliably constrain the peak energy \epo\ of their 
$\nu F_{\nu}$ spectrum by analyzing their time--integrated spectrum between 8 keV and 35 MeV.  
80\% of these spectra are fitted by a power--law with an exponential cutoff, 
and the remaining with the Band function.
Among these 318 GRBs, 274 and 44 belong to the long and short GRB class, respectively. 
Long GRBs have a typical peak energy \epo$\sim$160 keV and low energy spectral index $\alpha\sim -0.92$. 
Short GRBs have harder peak energy (\epo$\sim$490 keV) and harder low energy 
spectral index ($\alpha\sim -0.50$) than long bursts. 
For each \fe\ GRB we analyzed also the spectrum corresponding to the peak flux of the burst. 
On average, the peak spectrum has harder low energy spectral 
index but similar \epo\ than the corresponding time--integrated spectrum for the same burst.  
The spectral parameters derived in our analysis of \fe/GBM bursts are globally consistent with those reported 
in the GRB Cicular Network (GCN) archive after December 2008, 
while we found systematic differences, concerning the low energy power law index,
for earlier bursts. 
}
\maketitle

\section{Introduction}

Our current knowledge of the spectral properties of the prompt emission in GRBs mainly relies on the data 
collected in almost 10 years by the Burst And Transient Source Experiment (BATSE) onboard the 
{\it Compton Gamma--Ray Observatory (CGRO)}. 
\ba\ allowed to characterize the spectrum of the population of short and long GRBs over a large 
energy range, from 20 keV to 1--2 MeV. 
The analysis of such data revealed some important results about the spectral properties of these GRBs. 
The prompt spectra, integrated over the GRB duration (i.e. time integrated spectra), can be 
typically well described by a curved function showing a peak -- in a $\nu F_\nu$ representation -- at a 
typical energy \epo\ of a few hundreds of keV but whose distribution spans nearly three orders of magnitude. 
Large dispersions characterise also the distributions of the low--and high--energy 
photon indices, whose characteristic values are $\alpha\sim -1$ and 
$\beta\sim -2.3$, respectively (Band et al. 1993; Ghirlanda et al. 2002; Kaneko et al. 2006).
Similar results are obtained by considering the time resolved spectral analysis 
of flux/fluence limited samples of bright \ba\ bursts (Preece et al. 1998; Preece et al. 2000; Kaneko et al. 2006).

The BATSE data also suggested the existence of two different classes of GRBs (long and short), 
based on both temporal and spectral features. 
Evidence of a spectral diversity between long and 
short bursts comes from their different hardness--ratios (HR) (Kouveliotou et a. 1993).
The larger HR of short bursts might be ascribed to a larger \epo. Nava et al. (2008) and Ghirlanda et al. (2009 -- G09 hereafter) showed that \epo\ correlates both with the fluence and the peak flux. Although short and long bursts follow the very same \epo--peak flux relation, they obey different (parallel) \epo--fluence relations. This implies, obviously, that the distributions of the ratio \epo/fluence is different for the two burst classes.
Recently, Goldstein, Preece \& Briggs (2010) proposed this ratio as discriminator between short and long GRBs.
Due to the relation between \epo\ and the bolometric fluence and peak flux, a 
direct comparison between the \epo\ distributions of the two different burst classes must 
take into account the different fluence/peak flux selection criteria. G09 analyzed and compared samples of short 
and long BATSE bursts selected with similar peak flux limits. 
They found that the peak energy distributions of the two classes are similar, while 
the most significant difference concerns the low--energy power--law indices, with short 
bursts having typically a harder $\alpha\sim -0.4$.

These global spectral properties of GRBs have also been confirmed by other satellites
({\it Beppo}SAX, Hete--II and Swift) (Guidorzi et al. 2010; Sakamoto et al. 2005; Butler et al. 2007).
However, the detectors on board these satellites have different sensitivities with respect to 
\ba\ and cover a narrower and different energy range. 
For instance, the relatively narrow energy range (15--150 keV) of \sw/BAT does not 
allow to constrain the spectral peak \epo\ for most of the detected bursts (Cabrera et al. 2007; Butler et al. 2007).

Spectral studies of the prompt emission of GRBs require a wide energy range, 
possibly extending from few tens of keV to the MeV energy range. 
This allows to measure the curvature of the GRB spectrum and to 
constrain its peak energy, as well as its low and high energy spectral slopes. 

The \fe\ satellite, launched in June 2008, represents a powerful opportunity to shed light on 
the origin of the GRB prompt emission thanks to its two instruments: the Large Area Telescope (LAT) 
and the Gamma--ray Burst Monitor (GBM). 
LAT detected in about 2 years very high energy emission 
($>$ 100 MeV) from 19 GRBs. 
This emission component shares some common features with that already 
found in few bursts by EGRET (Energetic Gamma--Ray Experiment Telescope)
on board the {\it CGRO} satellite. 
In particular, high energy $\sim$GeV flux is still observed when 
the softer energy emission (in the sub--MeV domain) is ceased, and often its onset lags 
the sub--MeV component (e.g. Ghisellini et al. 2010, Omodei et al. 2009).

The other instrument on board \fe\ is the GBM (Meegan et al. 2009) which is similar to \ba\ and, 
despite its slightly worse sensitivity, allows to study the GRB spectrum 
over an unprecedented wide energy range, from 8 keV to 40 MeV. 
This is achieved by its twelve NaI detectors giving a good spectral 
resolution between $\sim$8 keV and $\sim$1 MeV and two BGO detectors which extend the energy range up 
to $\sim$40 MeV. 
Similarly to \ba, the NaI detectors guarantee full--sky coverage, but their smaller 
geometric area (16 times lower than that of the LADs of \ba) implies a lower sensitivity. 
On the other hand, the presence of the BGO detectors allows for the first time to extend the 
energy range for the study of the spectrum of the prompt emission to tens of MeV thus accessing 
an energy range poorly explored with the CGRO instruments.  

438 events, classified as GRBs\footnote{http://heasarc.gsfc.nasa.gov/W3Browse/fermi/fermigbrst.html}, 
triggered the GBM until the end of March 2010. 
{\it With this large sample of bursts we can perform, for the first time, 
a robust statistical study of the spectral properties of \fe/GBM bursts}. 
The main aims of this paper are: 
(a) to present the results of the spectral analysis of 438 \fe\  bursts, 
(b) to show the distribution of their spectral parameters, 
(c) to compare the spectral properties of \fe\ short and long GRBs and 
(d) to compare the spectra integrated over the burst duration 
(time--integrated spectra, hereafter) with the spectra of the most intense phase of the burst, 
i.e. its peak flux (peak spectra, hereafter). 

Preliminary results of the spectral analysis of \fe\ GRBs  performed by the GBM team have been 
distributed to the community through the Galactic Coordinates Network (GCN) Circulars. 
These amount to 228 GRBs  (until March 2010), whereof 167 have a well constrained \epo. 
On going spectral calibrations of the GBM detectors make the results published in the 
GCN  ``preliminary'', especially for the first bursts detected by the GBM. 
The GBM team continuously provides, together with the public data of detected GRBs, 
more updated detector response files. 
Few months ago the software and the new response files have been made public 
so that a systematic and reliable analysis of the spectra of \fe/GBM bursts is now possible. 
We will compare the results of our spectral analysis with those 
published in the GCN to search for possible systematic effects in the GCN results. 

The paper is organized as follows: in \S 2 and \S 3 we describe the sample of \fe\ GRBs 
and the procedure adopted to extract and analyze their spectral data, respectively. 
In \S 4 we present the spectral results and build their distributions, also considering 
short and long GRBs separately. 
In \S 4 we also compare the time--integrated spectra and the 
peak spectra for the analyzed bursts. 
We summarize our results in \S 5.

\section{The sample}

The GBM detected 438 GRBs up to the end of March 2010. 
A list of the GRB trigger number and the position in the sky, computed by the GBM, 
is provided by the Fermi Science Support Center$^1$.
The data of each GRB are archived and made public since July 2008.  
Note that, since the only information given in the public archive is the burst position, 
it is not possible to apply any selection in flux, fluence or duration on the GBM
on--line public archive since a spectral catalog is not available. 
For this reason we started the systematic analysis of GBM bursts in 
order to determine the spectral parameters, fluence and peak flux of all
bursts detected by \fe/GBM up to March 2010.

Preliminary spectral results of \fe\ GRBs have been distributed by the GBM team through the GCN system. 
We note that, starting on March 2010, the number of GCN of \fe\ bursts substantially decreased, although
the rate of detected bursts remained unchanged.
We decided to limit our selection to March 2010, thus 
having a large sample of GCN results to compare with.

We collected all bursts with spectral information published in the GCN up to the end of March 2010 
(228 objects). 
Among these 148 long GRBs and 19 
short have well constrained spectral parameters, in particular the peak energy of their $\nu F_{\nu}$. 
In \S 4 we will compare the GCN spectral parameters with those derived by us for the same bursts and search 
for possible systematic effects in the GCN results.

\section{Spectral analysis}
\label{models}
For the spectral analysis of \fe/GBM bursts we used the recently publicly 
released {\texttt rmfit - v3.3pr7} software\footnote{http://fermi.gsfc.nasa.gov/ssc/data/analysis/}.

For each GRB to be analyzed, the spectral analysis has been done by combining together more than one detector. Following the criterion adopted in Guiriec et al. (2010) and Ghirlanda et al. (2010), we selected the most illuminated NaI detectors having an angle between the source and the detector normal lower than 80 degrees. We selected the BGO \#0 or \#1 if the selected NaI were all between \#0-5 or \#6-12, respectively. When the selected NaI were both between \#0-5 and \#6-12, both BGO detectors were used. However, if one of the two BGO has zenith angle to the source larger than 100 degrees we exclude it from the analysis.

The very wide available energy range (from 8 keV to $\sim$40 MeV) allows to properly constrain 
the peak energy of particularly 
hard GRBs (with \epo\ larger than 1 MeV, i.e. the upper energy threshold of the NaI detectors) 
or the high energy spectral power law, if present. 

For the spectral analysis we used the CSPEC data (with time resolution of 1.024 s after the trigger time and 4.096 s before)
for long GRBs and the TTE data (with time resolution of 0.064 s) for short GRBs. A first hint about the burst duration comes from the visual inspection of the lightcurves (at different temporal resolutions) stored in the quicklook directory provided with the data$^1$. We used this method to decide what type of data (CSPEC or TTE) is more suitable for spectral analysis.
Both data type contain spectra with 128 energy channels. Following the prescription of the 
{\texttt rmfit} tutorial\footnote{http://fermi.gsfc.nasa.gov/ssc/data/analysis/user/vc\_rmfit\_tutorial.pdf} we considered 
the spectral data of the NaI detectors in the range 8 keV -- 900 keV and for the BGO detectors in the range 
250 keV -- 35 MeV.

For each GRB we extracted the background spectrum by selecting a time interval before and after the burst as large as possible, but distant enough from the burst signal, in order to avoid burst contamination. 
The spectra in these two time intervals were modeled in time with a polynomial of order between 
0 and 4 to account for the possible time evolution of the background. 
Then, the spectral analysis software extrapolates the background to the time interval occupied by the burst. 
We used the most updated response files with extension {\texttt rsp2}, which allows {\texttt rmfit} to use a new response for every 5 deg of spacecraft slew, as explained in the {\texttt rmfit} tutorial.

Each spectrum was analyzed adopting the Castor statistics (C--stat). 
Since we combined NaI and BGO detectors in the spectral analysis, we fitted the spectra 
allowing for a calibration constant among the different detectors. 
The spectral results (C--stat and spectral parameters) obtained with the calibration constants 
free and fixed to 1 were compared. 
If no significant difference was found between the C--stat and the spectral parameters obtained in these two cases, 
the calibration constants were fixed to 1 (as also suggested in the {\texttt rmfit} manual). In nearly 30\% of cases the C--stat significantly decreases by using free calibration constants. This is not directly related to the burst brightness (also for faint bursts the calibration constants can be required) even if, of course, the largest differences in C--stat values are found for bright bursts, since possible calibration offsets between instruments strongly affect the fit (in terms of C--stat) when data points have small errors.

Systematic residuals around the k-edge of the NaI detectors are often visible, owing to calibration issues (Guiriec et al. 2010). For 4 bursts (for which this effect is particularly pronounced) we performed the spectral analysis both including and excluding a few channels between 30 keV and 40 keV (e.g. see Guiriec et al. 2010). While the spectral parameters and their errors are not sensitive to this choice, the value of the C--stat is quite different. For these 4 bursts we report the results of both the analysis (Table 2).

\subsection{Spectral models}
The spectral analysis performed by different authors (Preece et al. 2000; Sakamoto et al. 2005; Kaneko et al. 2006; Butler et al. 2007; Nava et al. 2008; Guidorzi et al. 2010) on data taken from different 
instruments revealed that GRB spectra are 
 fitted by different models, the simplest ones being: 
i) a single power law model (PL), 
ii) a Band function (Band et al. 1993), which consists of two smoothly connected power laws and 
iii) a Comptonized model (CPL hereafter), i.e. a power law with a high energy exponential cutoff. 

The time--resolved spectra of BATSE GRBs have been also fitted combining thermal (black--body) and non thermal (power law) models (e. g. Ryde et al. 2005, Ryde et al. 2009). Although these fits are intriguing for their possible physical implications (Pe'er et al. 2010), they are statistically equivalent to fits with the phenomenological models described above (Ghirlanda et al. 2007). Recently, Guiriec et al. (2010) found evidence of a thermal black body component (summed to the standard Band function) in the spectrum of the Fermi GRB 100724B.

A simple PL function clearly indicates that no break/peak energy is detected within the energy range of the instrument. 
Furthermore, it is also statistically the best choice when the signal to noise of the analyzed spectra 
is very low because this model has the lowest number of free parameters. 
This was shown, for instance, by the analysis of the \sw/BAT spectral data (e.g. Cabrera et al. 2007).

The Band model (Band et al. 1993) has four free parameters to describe the low and high power law behaviours, 
the spectral break and the flux normalisation. 
Typically, the low energy photon index  $\alpha > -2$ [$N(E)\propto E^{\alpha}$] and the high energy 
photon index $\beta < -2$ [$N(E)\propto E^{\beta}$], so that a peak in $\nu F_\nu$ can be defined. 
When there is no evidence for a high energy photon tail or \epo\ 
is near the high energy boundary of the instrument sensitivity (and $\beta$ is poorly constrained) 
a CPL model is preferred, due to the lower number of parameters. Also in this case a peak energy 
can be defined when $\alpha>-2$. In the Band model the spectral curvature is fixed by $\alpha$, $\beta$ and \epo. 

We fitted all these models (which are all nested models) to each GRB spectrum. The addition of one free parameter requires an improvement in C-stat of 9 for a 3$\sigma$ confidence in this improvement. We chose this criterion to select the best fit model. In addition, we also require that all the spectral parameters are well determined (i.e. no upper or lower limits).

\subsection{Time integrated and peak flux spectra}

As anticipated we analyzed for each burst (1) the spectrum integrated over its whole duration and 
(2) the spectrum corresponding to the peak of the burst. 
Concerning the peak spectrum, this could be selected from the raw count light curve as 
the temporal bin with the largest count rate. 
However, it  may happen that bins with similar count rate have very different spectra 
and their flux can be considerably different. 
Therefore, a more physical approach for identifying the peak of the burst is to build 
its flux light curve (i.e. calculating the flux in physical units). 
In practice, we performed a time resolved spectral analysis of each burst and built its flux light curve 
(where the flux is integrated over the 8 keV--35 MeV energy range, i.e. the same spectral range where the spectral analysis is performed).
Then we identified the time bin (on the timescale of the data resolution, which is typically 1.024 and 0.064 
seconds for the long and short GRBs, 
respectively) corresponding to the largest flux and analyzed this spectrum to extract the peak 
spectrum parameters. 
We adopt this procedure for all GRBs, i.e. even those having a time integrated spectrum
better described by a simple power law. 

\section{Results}

The spectral parameters obtained from the analysis of the time--integrated spectra of the 438 \fe/GBM bursts 
are reported in Tab. \ref{tab1} and Tab. \ref{tab11}. 
In particular, in Tab. \ref{tab1} we list all the 323 bursts whose spectrum could be fitted with either 
the Band or CPL model (Col. 3). In five cases, the high-energy power-law index $\beta$ is $>-2$: this reduces the number of bursts with well defined peak energy \epo\ to 318. In Tab. \ref{tab11} we report all the 
109 cases where a single power law is the best fit to the data and the 6 cases where the spectral analysis was not possible due to lack of data.

In both tables we give the time interval over which the time--integrated spectrum 
was accumulated and the best fit model (Col. 2, 3), 
the normalization constant (Col. 4) in units of photons cm$^{-2}$ s$^{-1}$ (computed at 100 keV for all 
models) and the spectral index $\alpha$ (Col. 5) of the low energy power law. The peak energy of the 
$\nu F_{\nu}$ spectrum and (for the spectra fitted with the Band model) 
the high energy spectral index $\beta$ are listed in Col. 6 and Col. 7 of Tab. \ref{tab1}, respectively.
We also report in both tables the value of the C--stat resulting from the fit and the associated degrees 
of freedom (dof). 
The last column in Tab. \ref{tab1} gives the fluence obtained by integrating the best fit model over the 
8 keV--35 MeV energy range. 
For the spectra fitted with the PL model we give the fluence (last column in Tab. \ref{tab11}) computed 
over a smaller energy range, 8 keV--1 MeV, because we could not identify where the peak energy is. 
For 4 bursts (GRB 081009140, GRB 090618353, GRB 090626189, and GRB 090926181) we performed the spectral analysis both including and excluding a few channels around the k--edge. In these cases, we report in tables both the results. In 19 cases we found that the fit with a Band model returns well constrained parameters, but it is not statistically preferred to the Comp model (the C--stat improvement is lower than 9). In these cases we list in tables the parameters of both models.

In Tab. \ref{tab3} we report the results of the peak spectra analysis. In particular, we list the initial 
($t_1$) and final ($t_2$) time of the selected temporal bin, the best fit model, its spectral parameters, 
C--stat and degrees of freedom. The last column lists the peak flux estimated in the 8 keV--35 MeV energy range.

Finally, in Tab. \ref{tab4}, we list the spectral properties and the fluence collected from the 
GCN Circulars. For each bursts we also report the redshift (when available) and the GCN number.

\subsection{Time--integrated spectra}
Out of the 432 bursts for which it was possible to perform the spectral analysis, 359 are long and 73 are short. In the case of long (short) bursts, 274 (44) events have a well defined peak energy, while 4 (1) are best fitted with a Band model with $\beta>-2$. Most of the spectra are adequately fitted by the CPL model. This is true both for the long and short sub--groups. Among the 109 spectra fitted with a simple power law model there are 81 and 28 long and short events, respectively.  

In our analysis we integrated the spectrum over a time interval ($\Delta T$) where the signal of the burst (for all NaI detectors combined) is 
larger than the average background. Therefore, we adopt this integration time to separate short and long GRBs (i.e. $\Delta t < 2$s and $\Delta t > 2$ s, respectively). The distribution of $\Delta t$ is bimodal and short and long GBM bursts are well separated into two log normal distributions 
with central value (standard deviation) $< Log(\Delta t)> = 1.42$ ($\sigma$=0.39) and $< Log(\Delta t) > = -0.33$ ($\sigma$=0.38) for long and short GRBs, respectively.

\begin{figure}
\vskip -0.5 cm
\hskip -1.cm
\psfig{file=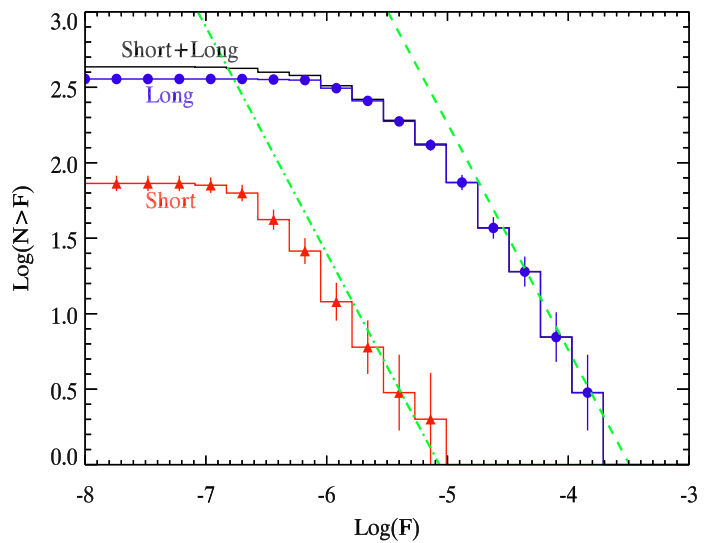,width=10.5cm,height=8cm}
\vskip -0.2 cm
\caption{Black histogram shows the $LogN-LogF$ of the 432 GRBs analyzed in this work (Tab. \ref{tab1} and \ref{tab11}).
Short GRBs (73 events) and long GRBs (359 events) are shown with (red) triangles and (blue) circles respectively. The dashed and dot--dashed lines
are two power laws with slope -3/2. The fluence $F$ in erg/cm$^2$ is obtained by integrating the best fit model in the 8 keV -- 1 MeV energy range. }
\label{fg0}
\end{figure}

\begin{figure}
\vskip -0.5 cm
\hskip -1.cm
\psfig{file=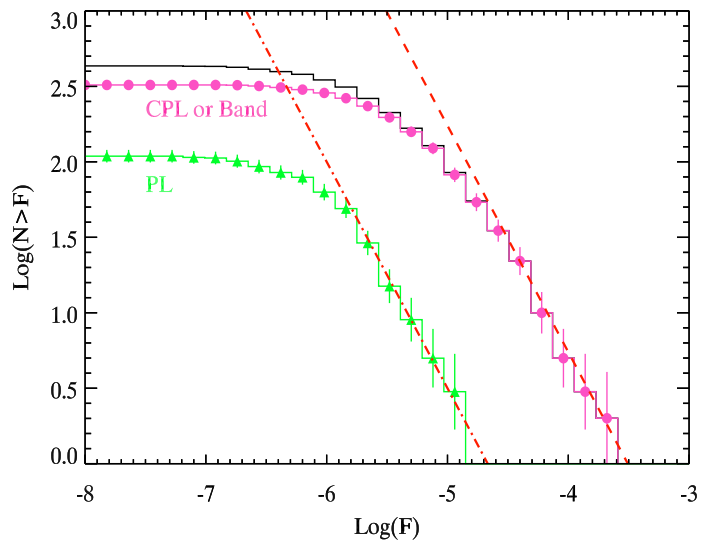,width=10.5cm,height=8cm}
\vskip -0.2 cm
\caption{$LogN-LogF$ distribution of all the GRBs fitted with the CPL or Band model 
and well constrained \epo\  (pink circles) and of the 109 GRBs fitted with a single PL 
model (green triangles). For reference a powerlaw with slope -3/2 is shown (dashed and dot--dashed line).
The fluence $F$ in erg/cm$^2$ is obtained by integrating the best fit model in the 8 keV -- 1 MeV energy range.
}
\label{fg00}
\end{figure}

{\it Fluence distribution ---} 
In Fig. \ref{fg0} we show the $LogN-LogF$  distributions of the \fe\ GRBs analyzed. 
In order to show the fluence distribution of all the 432 GRBs that we could successfully fit, 
we computed the fluence in the 8 keV--1 MeV for the 323 GRBs fitted with either the Band or CPL model 
and for the 109 GRBs fitted with a power law. We show separately the $LogN-LogF$ for 
long (359 events)  and short (73 events) GRBs.
At large fluences the distribution of long and short has a slope very similar to the euclidean one (-3/2), that is showed for comparison. In Fig. \ref{fg00} we show the $LogN-LogF$ distribution by dividing our sample according to the best fit model.

\begin{figure}
\vskip -0.5 cm
\hskip -1.15cm
\psfig{file=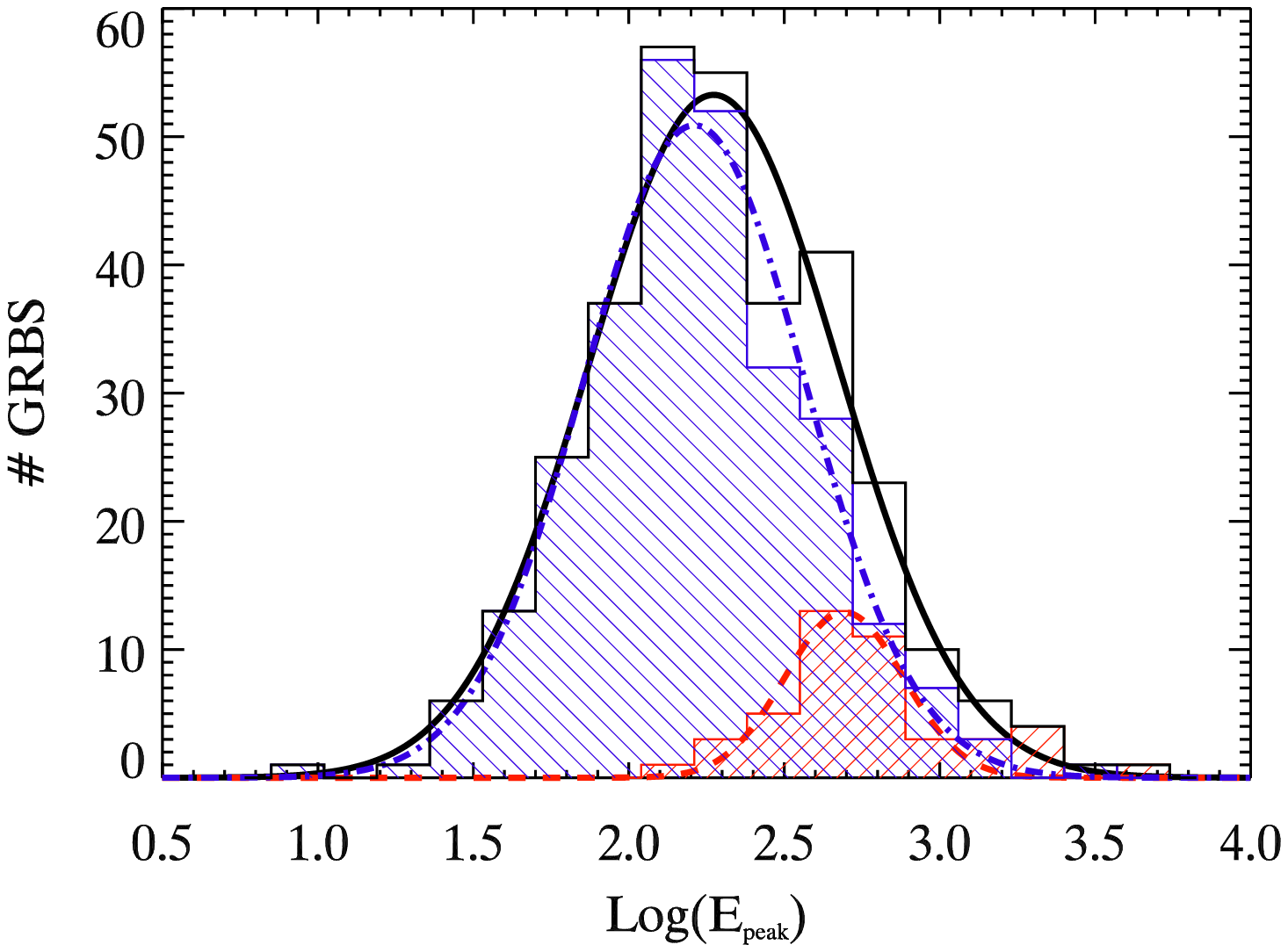,width=10.5cm,height=8cm}
\vskip -0.2 cm
\caption{
Distribution of the peak energy for the GRBs listed in Tab. \ref{tab1} fitted with either 
the Band or CPL model and with determined \epo\ (318 GRBs). 
The solid line shows the fit with a Gaussian. Also shown (hatched blue and red histograms) are the 
distributions for 274 long and 44 short GRBs, respectively, and their gaussian fits (dot--dashed and 
dashed line for long and short events, respectively).
}
\label{fg1}
\end{figure}

{\it $E_{\rm peak}$ distribution ---}
In Fig. \ref{fg1} we show the peak energy distribution of the 318 GRBs (both long and short) and 
the fit with a gaussian (solid black line). 
Also, in Fig. \ref{fg1} short and long events are shown separately and 
the Kolmogorov--Smirnov test gives a probability $P_{KS}=3.4\times 10^{-15}$ that the 
two distributions of \epo\ for long and short GRBs are drawn from the same parent distribution. 

\begin{figure}
\hskip -1.cm
\psfig{file=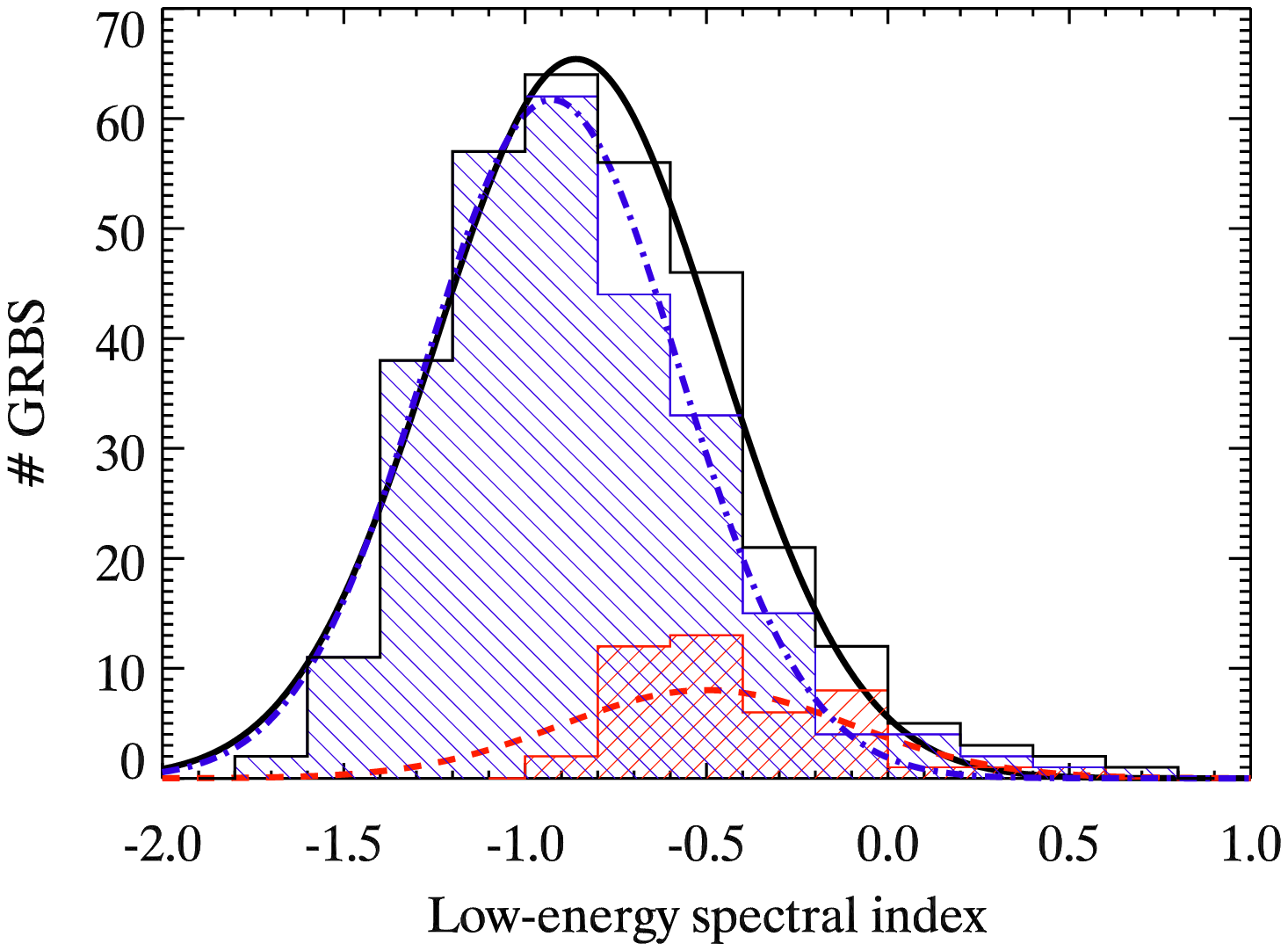,width=10.5cm,height=8cm}
\vskip -0.15 cm
\caption{
Distribution of the low-energy photon index for the 318 GRBs listed in Tab. \ref{tab1} 
fitted with either the Band or CPL model and with determined \epo. 
The solid (black) line shows the fit with a Gaussian. 
Also shown (hatched blue and red histograms) are the distributions for 274 long and 44 short GRBs, 
respectively, and their gaussian fits (dot--dashed and dashed line for long and short events, respectively). 
}
\label{fg2}
\end{figure}

\begin{figure}
\vskip -0.5 cm
\hskip -1.5cm
\psfig{file=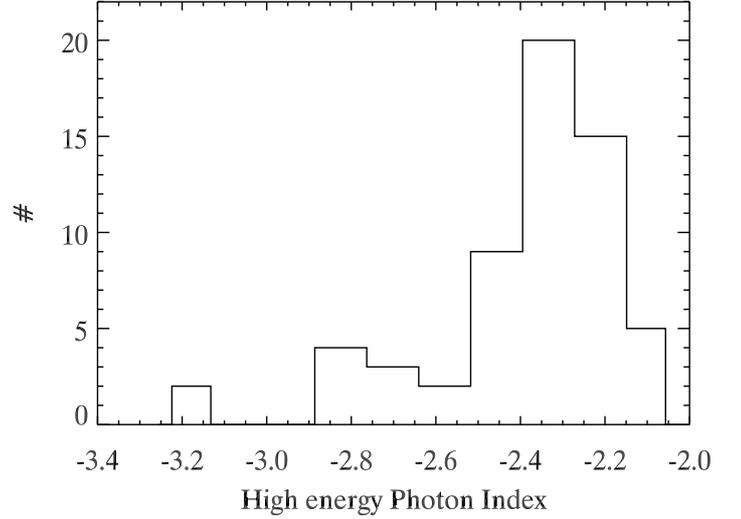,width=11cm,height=8cm}
\vskip -0.3 cm
\caption{
Distribution of the high-energy photon index for 60 GRBs whose time integrated spectrum 
is best fit with the Band model.}
\label{fg3}
\end{figure}

{\it Spectral index distributions ---}
In Fig. \ref{fg2} we show the distribution of the low energy spectral index $\alpha$ for all the 
318 GRBs and for short and long GRBs separately (having a $P_{KS}=7.3\times 10^{-12}$). 
Finally, in Fig. \ref{fg3} we show the distribution of the high-energy spectral index $\beta$ for 
the 60 time integrated spectra which are fitted with the Band model (see Tab. \ref{tab1}).

All the parameters distributions shown in Fig. \ref{fg1} and \ref{fg2} are fitted by 
gaussian functions whose parameters are reported in Tab. \ref{tab2}.

\begin{table}
\begin{center}
\begin{tabular}{lllcc}
\hline\hline
Parameter   &    Type & \# of GRBs   & Central value   &   $\sigma$    \\
\hline
\hline
Log(\epo)        & All     &318   &2.27		   &0.40          \\
			& Short &44   &2.69             &0.19		  \\
			& Long  &272  &2.21            &0.36            \\	   
\hline
$\alpha$     & All   &318  &--0.86		     &0.39          \\
             & Short &44   &--0.50             &0.40		  \\
             & Long  &274  &--0.92		     &0.35            \\	   
\hline	
\hline		
\end{tabular}
\caption{
Parameters of the gaussian fits of the distributions of the spectral parameters of 
\fe/GBM bursts analyzed in this work. 
We report the central value and the standard deviation considering all GRBs  (318 events),  short (44) and long (274) events with determined \epo.
}
\label{tab2}
\end{center}
\end{table}


\subsection{Peak Spectra}

As anticipated, for each burst we also extracted and analyzed the spectrum corresponding 
to its peak flux.
To this aim, we performed a time resolved spectral analysis of each burst. 
In the flux light curve we then selected the time bin (on the timescale given by 
the resolution of the data, i.e. typically 1.024 s and 0.064 s for long and short GRBs, 
respectively) with the largest flux.
We have also verified that in most cases the peak of the count rate coincides  
with the peak of the flux.

Out of 432 GRBs analyzed the peak spectrum could be extracted and fitted with a 
Band or CPL model in 235 cases. As before, the best fit model is defined by requiring an improvement in the C--stat value of 9 (between a given model and a more complex one with one parameter more), and well constrained spectral parameters. In 27 cases (26 long and 1 short) the best fit model of the peak spectrum is different from that of the time--integrated spectrum.
The spectral parameters of the peak spectra fitted with these two models are reported in 
Tab. \ref{tab3}. 

\begin{figure}
\vskip -0.3 cm
\psfig{file=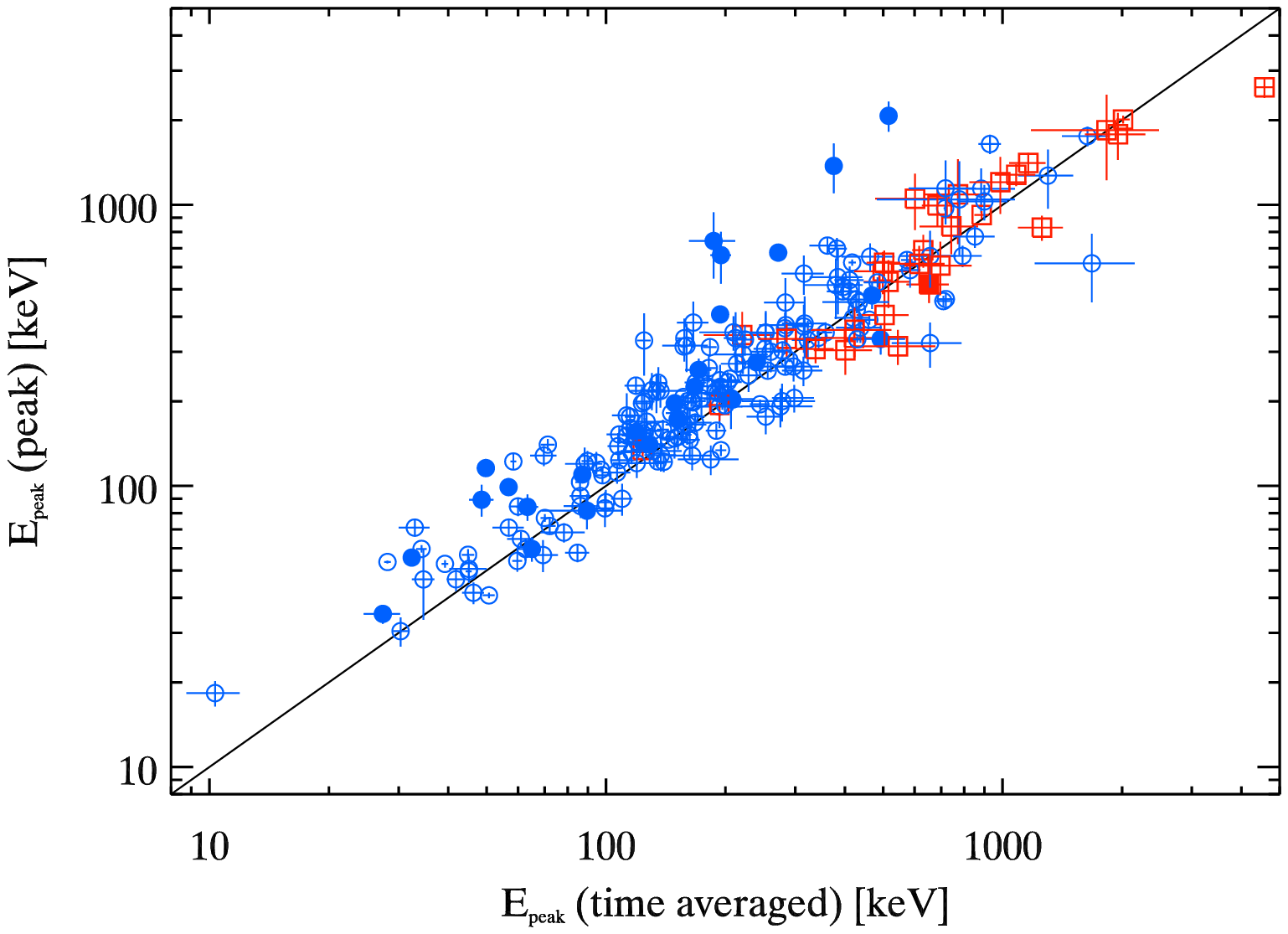,width=9cm,height=7cm}
\vskip -0.6 cm
\psfig{file=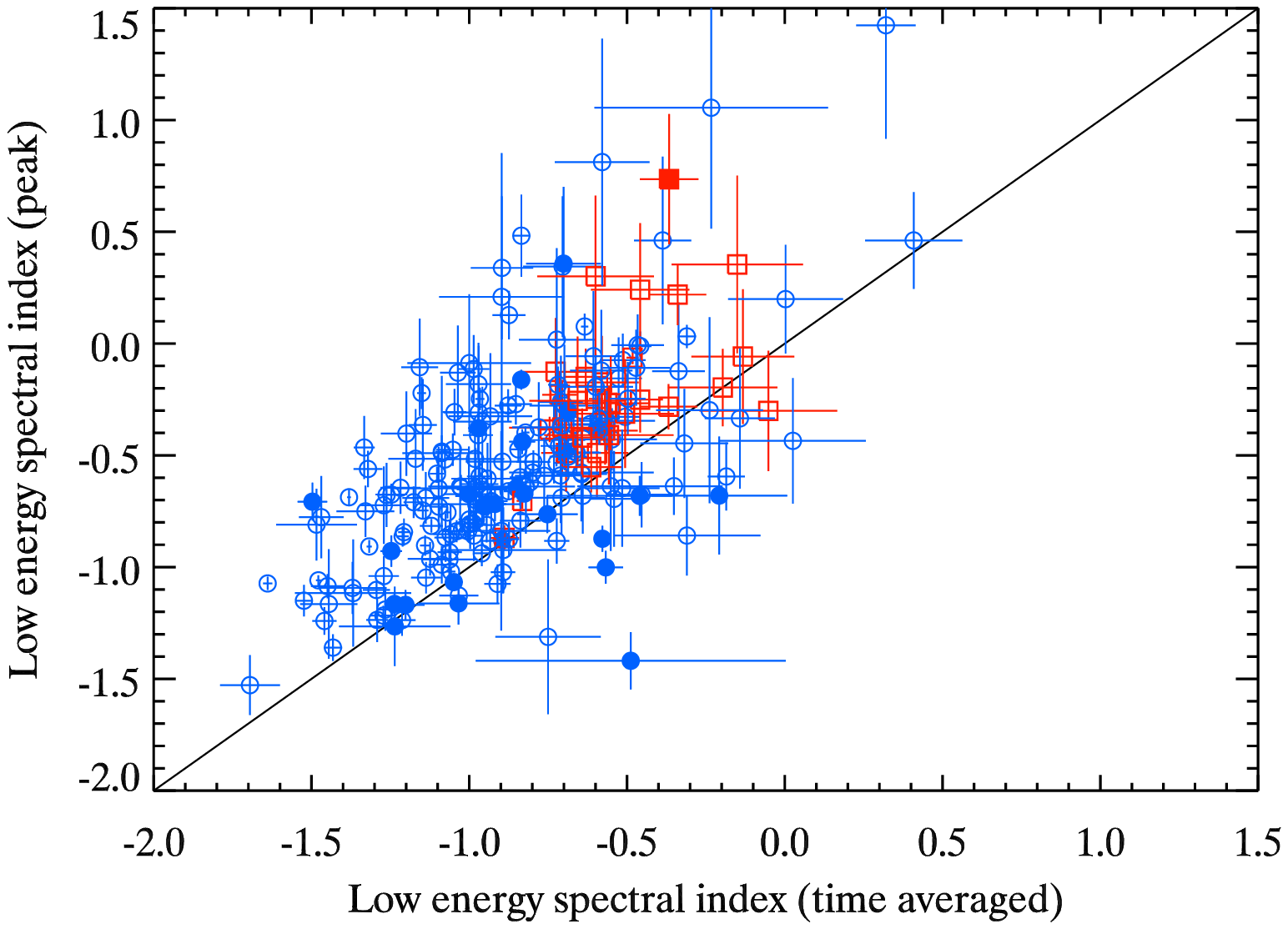,width=9cm,height=7cm}
\vskip -0.2 cm
\caption{
Comparison between time integrated and peak flux spectral parameters
for the 227 GRBs whose peak spectrum could be fitted with the Band or CPL model 
(reported in Tab. \ref{tab3} and present also in Tab. \ref{tab1}). 
Top panel: peak energy 
Bottom panel: low energy spectral index ($\alpha$).
Empty (filled) symbols are GRBs for which the time-integrated and the peak flux spectra have same (different) best fit model. Squares refer to short events and circles to long events.}
\label{fg5}
\end{figure}

In Fig. \ref{fg5} we compare the peak energy $E^{\rm obs}_{\rm peak}$ and the low spectral index $\alpha$ 
at the peak flux with the values of the time integrated spectrum for bursts having all these informations (227 events.).
Empty (filled) symbols refer to GRBs for which the time--integrated and the peak spectrum are described by the same (a different) best fit model. On average, the time integrated and peak spectrum values of $E^{\rm obs}_{\rm peak}$ are very similar, while the low energy spectral index, at the peak, is harder.

\begin{figure*} 
\includegraphics[scale=0.49]{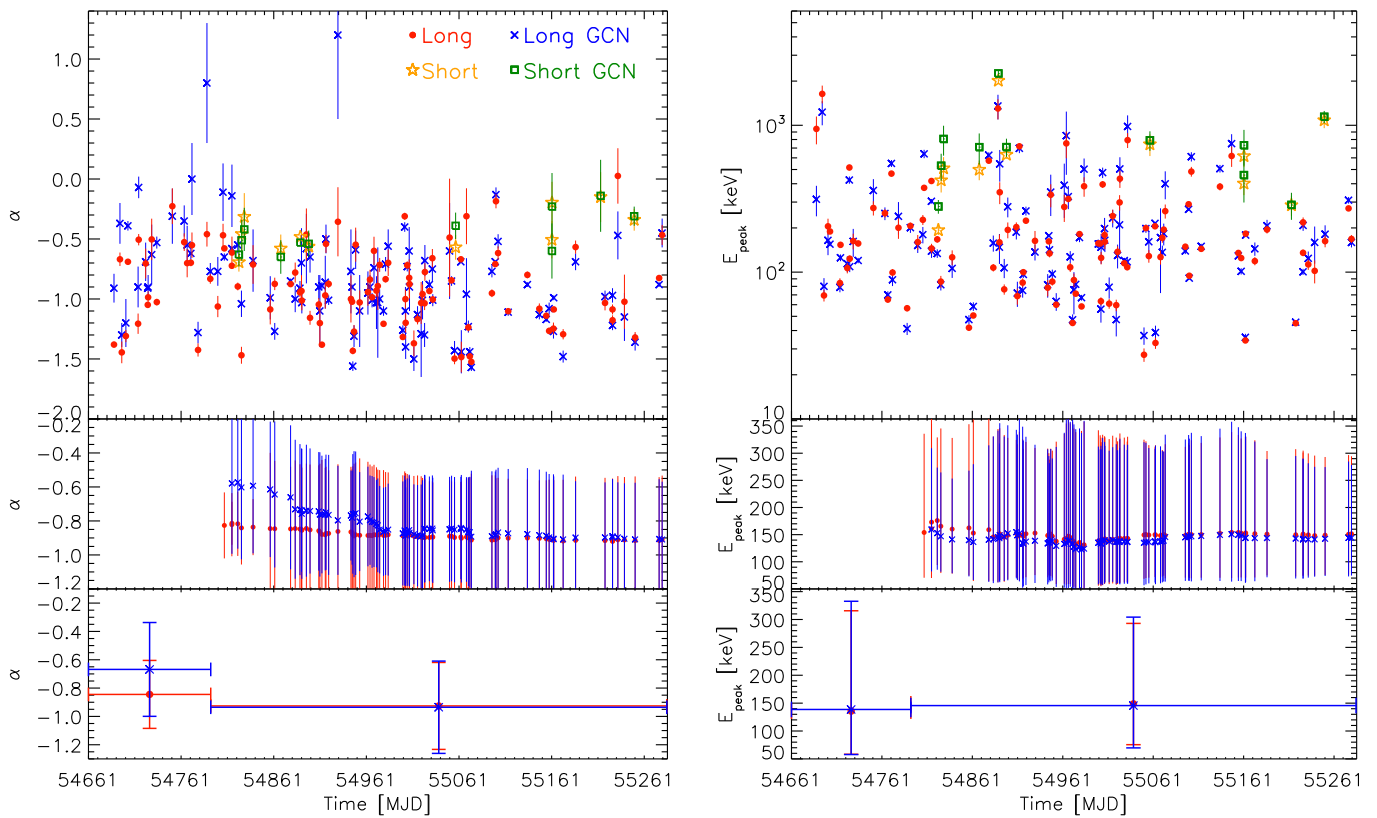}
\vskip -0.3 cm
\caption{
Upper panel: crosses and squares show (for long and short events respectively) 
the time trend of the spectral properties ($\alpha$ on the left and \epo\ on the right) 
for GBM bursts whose preliminary spectral analysis has been reported in the GCN Circulars. 
Circles and stars show the same for the sample of long and short bursts analyzed by us. We show only those bursts for which the same spectral models were used in our analysis and in the analysis reported in the GCN circulars.
Time on the x-axis is in MJD units. Middle panels: central values and 1$\sigma$ width 
for the $\alpha$ (left) and \epo\ (right) distributions (long bursts only) for the GCN sample 
(crosses) and our sample (circles) as a function of time.
Bottom panels show the average $\alpha$ and \epo\ for two different periods of time, till and after December 2008.}

\label{gcn} 
\end{figure*} 
\begin{figure} 
\includegraphics[scale=0.42]{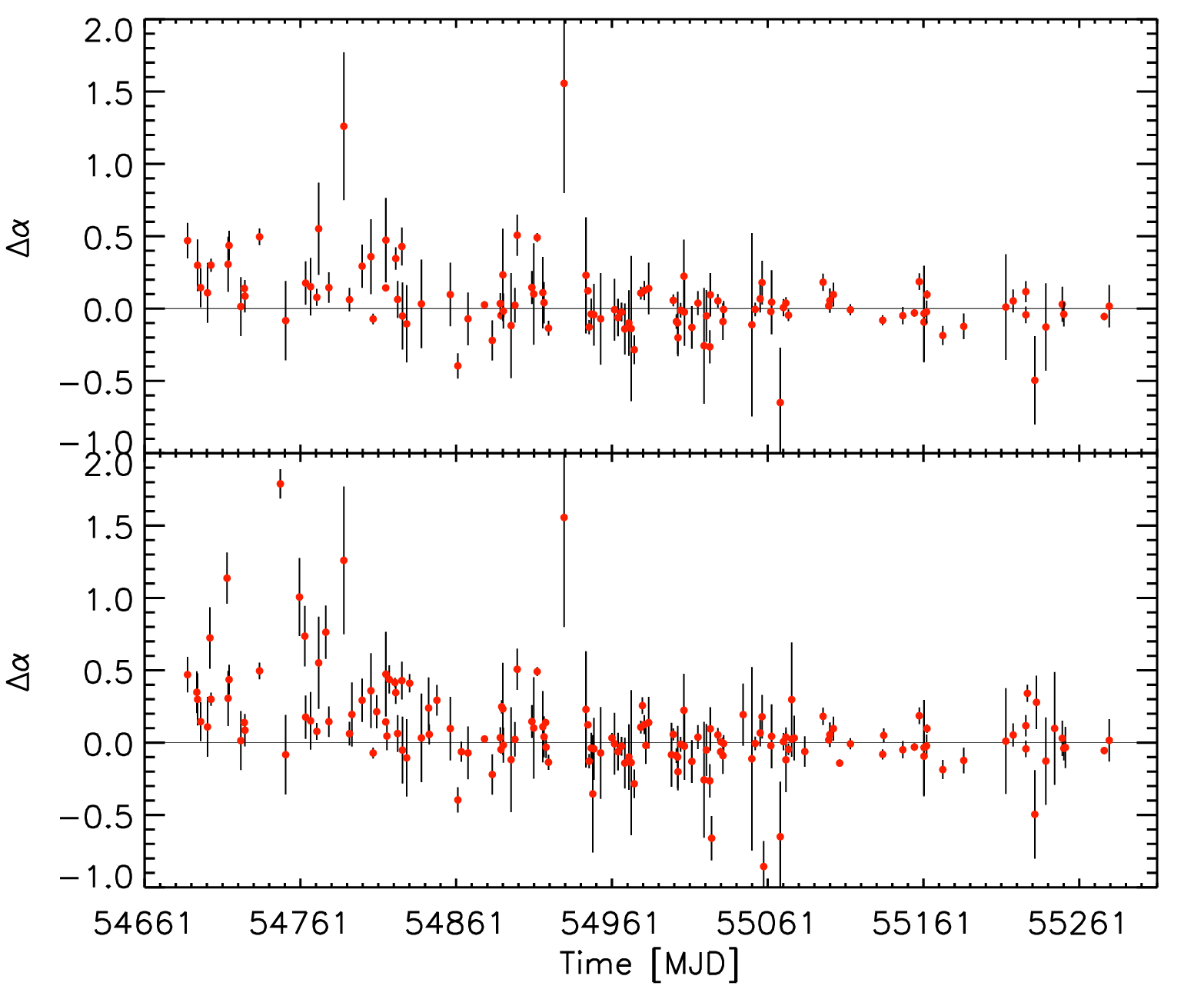}
\caption{Difference between $\alpha$ values reported in the GCN Circulars and $\alpha$ values 
derived from our analysis. Top panel: bursts for which the spectrum is described by the same model both in our analysis and in the GCN analysis. Bottom panel: bursts common to both samples no matter the spectral model chosen to describe the spectrum.}
\label{gcn2} 
\end{figure} 

\subsection{Comparison with GCN results}

Since August 2008, the Fermi GBM team is providing, through the GCN Circulars, 
preliminary results of the spectral analysis of a large number of the detected GRBs. 
For each burst, the GCN circular reports the burst's duration, spectral parameters, fluence and photon peak 
flux (all with their associated errors). GCN Circulars are promptly released when a burst occurs and are 
not updated since their first release. 
On the other hand, the GBM team is continuously providing, through the online archive, new versions of the detector 
response files, improved with respect to the first version used to perform their preliminary analysis. 
Our analysis benefits from the most updated response files. 
A meaningful comparison between our results and those preliminary reported by the GBM team 
must account for this difference. 

It is likely that the calibration of the different detectors changed and improved from the earliest 
to the latest Circular as well as the software and tools used by the Fermi team. 
If this is the case, spectral parameters of the first bursts detected by Fermi and reported in 
the GCN could be affected by systematic biases, hopefully not present in our analysis. 
To verify this possibility, we plotted the spectral parameters ($\alpha$ and \epo) as a 
function of the date (in MJD) of the GRB detection, to point out any possible systematic trend of their values 
and/or associated uncertainties. 
The spectral parameters (from our sample and from the GCN sample) are plotted in the 
upper panels of Fig. \ref{gcn}.

In the GCN sample, long bursts (crosses in Fig. \ref{gcn}) detected at the beginning of the 
mission have a slightly harder $\alpha$ with respect to the following bursts 
and also to the same bursts analysed by us (filled circles). 
This trend is not present in the short burst sample (squares in Fig. \ref{gcn}). 
Note, however, that in this case the sample is small and short bursts are present 
in the GCN sample only starting from December 2008 (empty squares).

A possible bias on the $\alpha$ values can be better quantified by probing how the $\alpha$ distribution 
for long bursts evolves in time. 
After having sorted GRBs according to their increasing date of detection, 
we consider the distribution of the spectral parameter up to a certain date and we fit it with a gaussian 
function, deriving its central value and the standard deviation. 
This corresponds to show how the {\it cumulative} distributions change by
increasing the time span. Since a systematic difference in $\alpha$ between our analysis and that reported in the GCNs might also be ascribed to a systematic difference in the choice of the best fit model, we use only those bursts for which the best fit model is the same in both the analysis.

Our test starts on November 2008, when we start to have enough bursts for a reliable gaussian fit. 
The middle left panel in Fig. \ref{gcn} shows our results. 
The y--axis shows the central value of the gaussian fit and 
the error bar correspond to its 1$\sigma$ width. 
In the GCN sample a trend is clearly visible. 
Although at the 1$\sigma$ level we see that all the central values are consistent, the mean of 
$\alpha$ (which is $\sim -0.5$ at the beginning) systematically evolves from harder to softer 
values and then it levels off at about --0.85. This result rules out the possibility that the bias is due to a different choice in the best fit model and suggests that another non physical effect could 
be biasing the low energy power law indices towards harder values in the first months of the GCN data analysis. 
This trend is not present in our sample (filled circles in Fig. \ref{gcn}, middle left panel), 
for which the $\alpha$ distribution does not evolve in time.

The whole $\alpha$ distributions for the two different samples are somewhat different, with GCN sample having slightly harder 
spectral indices ($\langle\alpha\rangle$=--0.86 with respect to $\langle\alpha\rangle$=--0.92 
derived from our analysis), as shown by the last point in the middle--left panel of Fig. \ref{gcn}.
We are interested in understanding if this difference can be totally ascribed to the bias affecting 
the first bursts and, in this case, in determining the date from which GCN preliminary results 
are consistent with those obtained by our analysis with the most updated response files. 
The bottom panels of Fig. \ref{gcn} shows the average values of $\alpha$ and \epo\ for two different periods of time: till the end of December 2008 and from January 2009  to March 2010, respectively.

Fig. \ref{gcn2} shows the difference $\Delta \alpha$ between the low energy spectral index as reported in the GCN 
sample and as derived by us, for bursts common to both samples. The upper panel shows bursts fitted with the same best fit model in both the analysis (i.e. the same sub--sample used in Fig. \ref{gcn}). The bottom panel, instead, shows all bursts common to both samples.
Approximately up to the end of 
December 2008 this difference is not randomly distributed around $\Delta\alpha=0$, but is systematically larger. This justifies our choice of considering these two time intervals for the 
bottom panels of Fig. \ref{gcn}, showing that bursts in the GCN sample up to the end of 
December 2008 have a mean $\alpha=-0.5$, while the $\alpha$ distribution of the remaining bursts is 
peaked around $\alpha=-0.9$, perfectly consistent with our results. 
The same separation has been applied to our sample, which does not show any difference in $\alpha$ 
when comparing bursts before and after December 2008 (bottom left panel in Fig. \ref{gcn}, filled circels).

The \epo\ values (right panels in Fig. \ref{gcn}) are untouched by this effect. 
The results from the two different samples are highly consistent. A weak trend is visible but 
note that the central value of the \epo\ distribution spans 
between 130 keV and 150 keV, a very small range if compared to the width of the distribution and 
to the typical errors on this parameter.

\section{Summary and Conclusions}

We analyzed the spectra of all GRBs detected by the \fe\ Gamma-ray Burst Monitor (GBM) between 
14 July 2008 and 30 March 2010. These are 438 GRBs and for 432 of them we have all the needed data to perform the 
spectral analysis. The time--integrated spectrum is best fitted with a power law model (110 spectra -- reported in Tab. \ref{tab11}) or a curved 
model (323 spectra -- reported in Tab. \ref{tab1})  which is either the Band model (65 spectra) or a 
cutoff--power law (CPL) model (258 spectra). 

Among the 432 GRBs for which we could analyze the spectrum, we identify 73 short and 359 long bursts, respectively. 
Their $LogN-LogF$ is similar (Fig. \ref{fg0}) and its high--fluence tail is consistent with a powerlaw with slope -3/2.

The 73\% of the bursts detected by the GBM up to March 2010 could be 
fitted with a curved model (Band or CPL, with a prevalence of the latter model) and in 
the majority (318 out of 323) of these cases we could constrain the spectral 
parameters and in particular the peak energy \epo\ of the $\nu F_{\nu}$ spectrum. 
This is possible thanks to the large energy range of the GBM 
spectra extending from 8 keV to $\sim$35 MeV.
This is the sample we considered for the characterization of the 
spectral parameters of the time--integrated spectra of \fe\ GRBs. 
Within this sample there are 44 short and 274 long GRBs. 
The comparison of their spectral properties 
shows that short GRBs have higher \epo\ than long events (Fig. \ref{fg1}) and a 
slightly harder low energy spectral index $\alpha$ (Fig. \ref{fg2}).

The finding that short \fe\ GRBs have harder peak energy than long events seems opposite 
to what found from the comparison of short and long GRBs detected by BATSE (Ghirlanda et al. 2009). 
However, the \fe\ short GRBs have also larger peak fluxes than long events. 
A more detailed comparison between long and short GRBs detected by {\it Fermi}/GBM and BATSE is presented in Nava et al. 2011.

A second major part of the present work was aimed to characterize the spectra of the peak of each GRB. 
Through time resolved spectroscopy we isolated and analyzed the spectrum corresponding to the peak 
of the flux light curve of each burst. The results are reported in Tab. \ref{tab3}.
By comparing the peak spectrum and the time--integrated spectrum of individual 
GRBs we find that the peak spectra have similar \epo\ of 
the time integrated spectra but harder low energy spectral index $\alpha$ (Fig. \ref{fg5}).

Finally we compared the results of our spectral analysis with those reported in the GCN circulars. 
We found that, due the still not fully completed calibrations of the GBM detectors, the GCN results 
of bursts comprised between July and December 2008 are affected by a systematic overestimate of 
the hardness of the GRB spectrum at low energies (i.e. the spectral parameter $\alpha$). 
This systematic bias does not affect \epo\ and is not present in our results which 
are obtained with the most recent releases of the GBM response files.

\begin{acknowledgements} 
This research has made use of the public \fe/GBM data and software obtained through the 
High Energy Astrophysics Science Archive Research Center Online Service, 
provided by the NASA/Goddard Space Flight Center. 
We acknowledge the GBM team for the public distribution of the spectral properties of 
\fe/GBM bursts through the GCN network. We thank the referee for his/her useful comments.
This work has been partly supported by ASI grant I/088/06/0.
LN thanks the Osservatorio Astronomico di Brera for the 
kind hospitality for the completion of this work. 
\end{acknowledgements}
\bibliographystyle{aa} 

\onecolumn
{\small
\longtab{2}{
\begin{longtable}{llllllllll}
\caption{Spectral parameters of the time averaged spectra of 323 \fe/GBM GRBs fitted with a curved function.  $^a$For all models the pivot energy for the amplitude is 100 keV. Fluences are in the range 8 keV -- 35 MeV.}\\

\hline\hline
GRB	      &$\Delta$t    &Model &$^a$A/$10^{-3}$  &$\alpha$  &$E_{\rm peak}$  &$\beta$  &C--stat  &dof  &Fluence/10$^{-7}$ \\
              &s                 &          &ph/cm$^2$/s         &               &keV                     &                  &         &       &erg/cm$^2$    \\
\hline
\\
\endfirsthead
\caption{continued. Spectral parameters of the time averaged spectra of 323 \fe/GBM GRBs fitted with a curved function. $^a$For all models the pivot energy for the amplitude is 100 keV. Fluences are in the range 8 keV -- 35 MeV.}\\
\hline\hline
GRB	      &$\Delta$t    &Model &$^a$A/$10^{-3}$  &$\alpha$  &$E_{\rm peak}$  &$\beta$  &C--stat  &dof  &Fluence/10$^{-7}$ \\
              &s                 &          &ph/cm$^2$/s         &               &keV                     &                  &         &       &erg/cm$^2$    \\
\hline
\\
\endhead
\hline
\endfoot
080714745 &   28.672 & CPL &    8.79$\pm$   1.41 &   --1.14$\pm$   0.07 &    167.7$\pm$    16.7 &    ... &    765.53 &          595 &   73.75$\pm$   3.73 \\
080715950 &   14.336 & CPL &    7.57$\pm$   0.48 &   --1.27$\pm$   0.04 &    315.5$\pm$    38.4 &    ... &    784.04 &          712 &   59.74$\pm$   3.16 \\
080717543 &   26.624 & Band &   10.92$\pm$   3.55 &   --0.66$\pm$   0.18 &    119.1$\pm$    19.8 &     --2.11$\pm$   0.17 &    740.76 &          462 &   85.89$\pm$  19.96 \\
080719529 &   44.032 & CPL &    4.67$\pm$   2.50 &   --0.56$\pm$   0.30 &     92.56$\pm$    15.0 &    ... &    450.01 &          348 &   14.87$\pm$   1.81 \\
080723557 &   73.728 & Band &   34.02$\pm$   0.72 &   --0.85$\pm$   0.01 &    194.0$\pm$     3.7 &     --2.43$\pm$   0.05 &   1281.70 &          468 & 1011.00$\pm$  32.41 \\
080723913 &    0.32 & CPL &   38.28$\pm$  14.60 &   --0.24$\pm$   0.30 &    168.7$\pm$    26.3 &    ... &    375.81 &          355 &    1.66$\pm$   0.20 \\
080723985 &   60.417 & CPL &   10.18$\pm$   0.24 &   --0.98$\pm$   0.02 &    427.4$\pm$    18.7 &    ... &    597.02 &          472 &  412.89$\pm$   9.67 \\
080724401 &  120.834 & Band &    9.50$\pm$   0.86 &   --0.97$\pm$   0.05 &     97.96$\pm$     4.86 &     --2.33$\pm$   0.09 &    889.96 &          591 &  270.22$\pm$  20.53 \\
\end{longtable}
}}

\twocolumn

\onecolumn

{\small
\longtab{3}{
\begin{longtable}{llllllllll}
\caption{Spectral parameters of the time--averaged spectra of 109 \fe/GBM GRBs fitted with a powerlaw. The fluence is computed in the 8 keV--1 MeV energy range. The 6 GRBs for which the analysis can not be performed (due to lack of data) are also listed. $^a$The pivot energy for the amplitude is 100 keV.}\\
\hline\hline
GRB	      &$\Delta$t    &Model &$^a$A/$10^{-3}$  &$\alpha$  &C--stat  &dof  &Fluence/10$^{-7}$ \\
                      &s                &           &ph/cm$^2$/s   &             &              &        &erg/cm$^2$    \\
\hline
\\
\endfirsthead
\caption{continued. Spectral parameters of the time--averaged spectra of 109 \fe/GBM GRBs fitted with a powerlaw. The fluence is computed in the 8 keV--1 MeV energy range. The 6 GRBs for which the analysis can not be performed (due to lack of data) are also listed. $^a$The pivot energy for the amplitude is 100 keV.}\\
\hline\hline
GRB	      &$\Delta$t    &Model &$^a$A/$10^{-3}$  &$\alpha$  &C--stat  &dof  &Fluence/10$^{-7}$ \\
                      &s                &           &ph/cm$^2$/s   &             &              &        &erg/cm$^2$    \\
\hline
\\
\endhead
\hline
\endfoot
080714086 &    7.17 & PL &    1.41$\pm$   0.11 &   --1.36$\pm$   0.04 &    522.07 &    475 &           10.45$\pm$     1.98 \\
080714425 &   41.98 & PL &    0.97$\pm$   0.06 &   --1.70$\pm$   0.05 &    338.38 &    352 &           33.09$\pm$     6.62 \\
080723339 &    ... & ... &    ... &    ... &     ... &     ... &           ... \\
080803772 &  121.86 & PL &    1.12$\pm$   0.03 &   --1.73$\pm$   0.02 &    605.62 &    356 &          115.4$\pm$    9.54 \\
080805496 &   38.91 & PL &    0.89$\pm$   0.05 &   --2.14$\pm$   0.04 &    583.33 &    474 &           27.55$\pm$     2.32 \\
080808451 &   23.55 & PL &    1.02$\pm$   0.09 &   --1.69$\pm$   0.06 &    539.85 &    470 &           19.73$\pm$     8.46 \\
080815917 &    1.09 & PL &    6.69$\pm$   0.37 &   --1.48$\pm$   0.04 &    474.87 &    359 &            6.80$\pm$     0.49 \\
080817720 &    7.17 & PL &    1.69$\pm$   0.18 &   --1.15$\pm$   0.04 &    441.30 &    480 &           15.96$\pm$    26.59 \\
\end{longtable}
}}

\twocolumn

 \onecolumn
{\scriptsize
\longtab{4}{
\begin{longtable}{llllllllllll}
\caption{Spectral parameters of the peak spectra of 235 \fe/GBM GRBs fitted with a curved function. $^a$For all models the pivot energy for the amplitude is 100 keV. The flux is computed in the 8 keV--35 MeV energy range.}\\
\hline\hline
GRB	     & $\Delta t$ & $t_1$ & $t_2$  & Model & $^a$A/10$^{-3}$ & $\alpha$   & $E_{\rm peak}$  & $\beta$  & C--stat  & dof  & Flux/10$^{-7}$ \\
             &  s          &    s    & s      &       & ph/cm$^2$/s &            & keV             &          &          &      &   erg/cm$^2$/s    \\
\hline
\\
\endfirsthead
\caption{continued. Spectral parameters of the peak spectra of 235 \fe/GBM GRBs fitted with a curved function. $^a$For all models the pivot energy for the amplitude is 100 keV. The flux is computed in the 8 keV--35 MeV energy range.}\\
\hline\hline
GRB	     & $\Delta t$ & $t_1$ & $t_2$  & Model & $^a$A/10$^{-3}$ & $\alpha$   & $E_{\rm peak}$  & $\beta$  & C--stat  & dof  & Flux/10$^{-7}$ \\
             &  s          &    s    & s      &       & ph/cm$^2$/s &            & keV             &          &          &      &   erg/cm$^2$/s    \\
\hline
\\
\endhead
\hline
\endfoot
080714745 & 28.672&	2.56 &   3.58 & Band & 752.90$\pm$712.00 &   0.72$\pm$  0.51 &    78.5$\pm$   10.5 &  --2.07$\pm$  0.08 &   669.28 &  	598 &  24.01$\pm$  3.80 \\
080715950 & 14.336&	1.02 &   2.05 & CPL &  21.67$\pm$  1.15 &  --1.22$\pm$  0.04 &   569.1$\pm$   90.2 & ... &   788.22 & 	 717 &  18.81$\pm$  1.50 \\
080723557 & 73.7288&  55.04 &  56.06 & CPL & 115.00$\pm$  3.05 &  --0.65$\pm$  0.03 &   407.3$\pm$   12.9 & ... &   522.12 & 	 472 &  72.47$\pm$  1.60 \\
080723985 & 60.417&  13.06 &  14.08 & CPL &  36.28$\pm$  1.87 &  --0.79$\pm$  0.05 &   460.2$\pm$   35.8 & ... &   523.09 & 	 475 &  26.58$\pm$  1.30 \\
080724401 & 120.834&  21.12 &  22.14 & Band & 132.70$\pm$ 23.80 &  --0.41$\pm$  0.10 &   109.1$\pm$	 7.12 &  --2.43$\pm$  0.12 &   661.31 & 	 595 &  17.25$\pm$  1.80 \\
080725435 & 37.888&	6.14 &   7.17 & CPL &  13.40$\pm$  1.77 &  --0.87$\pm$  0.11 &   330.3$\pm$   56.2 & ... &   501.36 & 	 477 &   6.67$\pm$  0.70 \\
080727964 & 81.921&  14.34 &  15.36 & CPL &  20.01$\pm$  2.84 &  --0.55$\pm$  0.13 &   286.1$\pm$   36.0 & ... &   947.11 & 	 829 &   7.54$\pm$  0.67 \\
080730520 & 24.576&	3.84 &   4.86 & CPL &  34.86$\pm$  6.16 &  --0.78$\pm$  0.12 &   148.6$\pm$   14.2 & ... &   756.82 & 	 699 &   6.26$\pm$  0.38 \\
\end{longtable}
}}

\twocolumn

 \onecolumn

{\small
\longtab{5}{
\begin{longtable}{cllllllll}
\caption{Spectral parameters of the \fe/GBM GRBs collected from the GCN Circular Archive. $^*$References in brackets are for the redshift.
(a) Greiner J., Clemens C., Kruhler T. et al., 2009, A\&A, 498, 89; (b) R. Salvaterra et al, 2009, Nature, 461, 1258.}\\
\hline\hline
GRB	      &   $z$ & $\Delta$t  & $E_{\rm peak}$  & $\alpha$ & $\beta$  & Fluence & Range & GCN$^*$ \\
		  &   &   s            & keV             &          &          & erg/cm$^2$   &keV & number\\
\hline
\\
\endfirsthead
\caption{continued. Spectral parameters of the \fe/GBM GRBs collected from the GCN Circular Archive. $^*$References in brackets are for the redshift.
(a) Greiner J., Clemens C., Kruhler T. et al., 2009, A\&A, 498, 89; (b) R. Salvaterra et al, 2009, Nature, 461, 1258.}\\
\hline\hline
GRB	      &   $z$ & $\Delta$t  & $E_{\rm peak}$  & $\alpha$ & $\beta$  & Fluence & Range & GCN$^*$ \\
		  &   &   s            & keV             &          &          & erg/cm$^2$   &keV & number\\
\hline
\\
\endhead
\hline
\endfoot
080810549 & 3.35  &  50    & 	313.5	$\pm$73.6  &   -0.91  $\pm$ 0.12  & ...  &  6.9e-6   $\pm$  0.5e-6   &  50-300	 &  (8083)8100    \\
080816503 &...&  70    & 	146.7	$\pm$14.5  &   -0.57  $\pm$ 0.14  & ...  &  1.86e-5  $\pm$  0.20e-5  & 50-300	 &  8107	    \\
080816989 &...&  5     & 	1230	$\pm$230   &   -0.37  $\pm$ 0.17  & ...  &  7.23e-7  $\pm$  4.2e-8   &  25-1000	 &  8111	    \\
080817720 &...&  6     & 	...     &   -1.07  $\pm$ 0.04  & ...  &  2.6e-6   $\pm$  0	 &    25-1000	 &  8109    \\
080818579 &...&  50    & 	...     &   -1.57  $\pm$ 0.05  & ...  &  2.26e-6  $\pm$  0	 &    50-300	 &  8110	    \\
080818945 &...&  10    & 	80	$\pm$11    &   -1.3   $\pm$ 0.1   & ...  &  1.e-6	 $\pm$  0	 &    50-300	 &  8116	    \\
080823363 &...&  46    & 	164.7	$\pm$34.2  &   -1.2   $\pm$ 0.2   & ...  &  4.1e-6   $\pm$  0	 &    50-300	 &  8145    \\
080824909 &...&  28    & 	100	$\pm$16    &   -0.4   $\pm$ 0.2   &  -2.1 $\pm$ 0.2  &  2.3e-6   $\pm$  0	 &    50-300	 &  8144    \\
\end{longtable}
}
}
\twocolumn

\end{document}